\documentclass[pdflatex,sn-mathphys-num]{sn-jnl}

\usepackage{graphicx}%
\usepackage{multirow}%
\usepackage{amsmath,amssymb,amsfonts}%
\usepackage{amsthm}%
\usepackage{mathrsfs}%
\usepackage[title]{appendix}%
\usepackage{xcolor}%
\usepackage{textcomp}%
\usepackage{manyfoot}%
\usepackage{booktabs}%
\usepackage{algorithm}%
\usepackage{algorithmicx}%
\usepackage{algpseudocode}%
\usepackage{listings}%
\usepackage{threeparttable}
\usepackage{url}
\usepackage{hyperref}
\usepackage{subcaption}
\usepackage[font=small,labelfont=bf]{caption}
\theoremstyle{thmstyleone}%
\theoremstyle{thmstyletwo}%
\theoremstyle{thmstylethree}%

\begin{document}

\title[Article Title]{A Resource-Aligned Hybrid Quantum-Classical Framework for Multimodal Face Anti-Spoofing}

\author[1]{\fnm{Wanqi} \sur{Sun}}\email{sunwanqi19@mails.ucas.ac.cn}
\author*[2]{\fnm{Jungang} \sur{Xu}}\email{xujg@ucas.ac.cn}
\author[1]{\fnm{Chenghua} \sur{Duan}}\email{cduan@ucas.ac.cn}

\affil[1]{\orgdiv{School of Electronic, Electrical and Communication Engineering}, \orgname{University of Chinese Academy of Sciences}, \orgaddress{\city{Beijing}, \country{China}, \postcode{100000}}}
\affil*[2]{\orgdiv{School of Computer Science and Technology}, \orgname{University of Chinese  Academy of Sciences}, \orgaddress{\city{Beijing}, \country{China}, \postcode{100000}}}

\abstract{
Embedding high-dimensional data into resource-limited quantum devices remains a significant challenge for practical quantum machine learning. In multimodal face anti-spoofing, while linear compression methods such as principal component analysis can reduce dimensionality to accommodate limited quantum budgets, such approaches often lose critical high-order cross-modal correlations due to the loss of structural information.
To this end, we propose a hybrid Matrix Product State (MPS)–Variational Quantum Circuit (VQC) framework, where the MPS serves as a structured, differentiable pre-quantum compression and fusion module, and the VQC acts as the quantum classifier. 
Built upon the low-rank structure controlled by the virtual bond dimension and integrated with a configurable nonlinear enhancement mechanism, this MPS module explicitly models long-range cross-modal correlations while compressing multimodal data into a compact representation matching the quantum budget and improving numerical stability under extreme compression.
Experiments on the CASIA-SURF benchmark demonstrate that MPS-VQC achieves accuracy comparable to strong classical neural network baselines with fewer than 0.25M parameters, highlighting the parameter efficiency of tensor-network representations for high-dimensional multimodal data under tight resource budgets.
Leveraging the intrinsic compatibility between MPS structures and quantum circuit topology, this framework not only provides a viable technological pathway for efficient multimodal anti-spoofing on NISQ devices but also serves as a stepping stone toward fully quantum implementations of such tasks in the future.
}



\maketitle

\section{Introduction}\label{sec1}
In high-security applications such as identity verification, mobile payment, and access control, face recognition systems remain vulnerable to presentation attacks \cite{yu2023deep, antil2025unmasking}. As a result, face anti-spoofing (FAS) has become a critical component for ensuring system reliability \cite{handbook2022, iso30107}. Compared with unimodal methods that rely only on visible-light images, multimodal FAS leverages complementary sensing channels, including RGB, depth, and infrared (IR), to capture texture cues, 3D shape, and imaging-related properties \cite{george2020biometric}. Studies on public benchmarks such as CASIA-SURF \cite{zhang2019dataset} show that multimodal data is more robust under challenging conditions, including complex lighting, occlusion, and diverse attack materials \cite{shen2019facebagnet}.
However, in real-world deployment, multimodal systems are constrained by tight computational and latency budgets \cite{ameenulhakeem2025lightweight, he2023lightweight}. They also often suffer from missing or degraded modalities due to sensor failures, cross-sensor misalignment, or environmental interference \cite{lin2024suppress, yu2025visual}. These practical issues make efficient and robust cross-modal fusion an urgent problem \cite{baltrusaitis2019multimodal}.

Existing multimodal FAS methods typically adopt early fusion, late fusion, or intermediate fusion. As summarized by Baltrusaitis et al.~\cite{baltrusaitis2019multimodal}, this involves a set of well-known trade-offs: earlier fusion enables stronger cross-modal interactions but is more sensitive to misalignment and noise, whereas later fusion is usually more robust but cannot fully exploit complementary information across modalities~\cite{neverova2016moddrop}. 
Although methods such as FaceBagNet introduce a bag-of-local-features strategy and improve robustness to modality missingness through mechanisms like random modality feature erasing~\cite{shen2019facebagnet, brendel2019approximating}, they often rely on high-dimensional feature representations. This fuse-then-compress paradigm---building cross-modal interactions in a high-dimensional space and then applying linear dimensionality reduction via fully connected layers or pooling---tends to discard high-order cross-modal interactions under tight dimension budgets~\cite{liu2018efficient}. As deep models scale up, this approach further increases the parameter count and computational cost, making it difficult to fit lightweight models that are highly sensitive to input dimensionality~\cite{zhang2025lightweight}. Therefore, we need a representation learning mechanism that unifies multimodal fusion and aggressive compression, while keeping model complexity under control.

Variational quantum circuits (VQCs) leverage the high-dimensional Hilbert space as a feature space and have been extensively explored for classification and feature extraction \cite{havlicek2019supervised, schuld2019quantum}. Theoretically, VQCs can use quantum entanglement to capture data features that are difficult for classical models \cite{huang2021power}. In practice, however, they are tightly constrained by physical resources, since the number of high-fidelity qubits and the available coherence time bound the achievable circuit width and depth \cite{bharti2022noisy}. Encoding high-dimensional features into a small quantum system leads to a trade-off between expressivity and trainability. Shallow circuits lack the capacity to capture discriminative structure, whereas deeper and more entangled circuits are more likely to suffer from barren plateaus during training \cite{mcclean2018barren}. 
Moreover, unavoidable noise in current quantum hardware accumulates as circuit depth increases. This further exacerbates the training challenges discussed above, making it difficult in practice to rely on deeper circuits to accommodate high-dimensional inputs \cite{wang2021noise, cerezo2022challenges}. Although some studies incorporate classical residual connections by concatenating parallel classical features at the readout stage to compensate for the limited expressivity of the quantum model \cite{liang2021hybrid}, this strategy effectively bypasses the bottlenecks of the quantum circuit. It does not resolve the central challenge of how to model the complex structure of high-dimensional data under strict quantum resource constraints. Consequently, achieving effective FAS on NISQ hardware hinges on a structured pre-quantum compression of multimodal information into a low-dimensional subspace that matches the available qubit budget, while preserving the discriminative features needed to distinguish live samples from spoofing attacks.

Tensor Networks (TN), particularly Matrix Product States, provide a theoretical foundation for mapping high-dimensional classical features to small-scale quantum systems due to their natural mathematical alignment with many-body quantum states~\cite{stoudenmire2016supervised, huggins2019towards}. 
First, MPS offers a representation with controlled complexity via low-rank decomposition. The core idea is to decompose a high-order global tensor into a contraction of local low-order tensors, where the expressivity is explicitly parameterized by the tensor train rank or virtual bond dimension~\cite{oseledets2011tensor, novikov2015tensorizing}. 
Second, modern TN applications have evolved from simple parameter compression into trainable interaction operators. They can explicitly approximate high-order couplings between multimodal features, enabling feature fusion and dimensionality reduction to occur simultaneously under the same low-rank constraint~\cite{jang2025tensor}. This efficiently captures key correlations across modalities without the need to construct and process a massive intermediate feature space~\cite{liu2018efficient}. These allow us to flexibly adjust the  dimension and complexity to match the qubit budget of the subsequent VQC.

To address these limitations, we propose MPS-VQC, a hybrid quantum-classical multimodal face anti-spoofing framework designed for the resource constraints of NISQ devices. The framework utilizes an MPS that operates as a structured, pre-quantum compression and fusion module, and a VQC that is employed as the quantum classifier.
Our main contributions are summarized as follows:
\begin{enumerate}
\item Structured, pre-quantum MPS fusion-compression module. This module performs structured fusion and efficient dimensionality reduction by leveraging the MPS's virtual bond indices to transmit and integrate information along its chain structure, thereby providing a compact representation for quantum state encoding in the subsequent VQC. During this process, MPS's parameter count is effectively reduced through virtual bond dimension truncation.

\item Numerically stable training with nonlinear enhancement. To address numerical instability in long-chain tensor contractions, we propose a dynamic update strategy that combines nonlinear activations with the mixed canonical form. This design mitigates gradient oscillations and rank collapse, improving training stability and convergence on long sequential inputs.

\item Two-stage decoupled optimization. Because long MPS chains can be sensitive to data variations and joint updates may introduce a non-stationary input distribution, we develop a time-decoupled, stepwise training scheme. It optimizes structured feature extraction and quantum classification in separate stages, reducing drastic feature-space drift.

\item Empirical evaluation with high parameter efficiency. Experiments on CASIA-SURF show that MPS-VQC achieves accuracy comparable to strong classical neural network baselines with only about 0.25M parameters, demonstrating the efficiency of tensor-network representations for high-dimensional multimodal data under tight resource budgets.

\end{enumerate}
In addition, since the MPS structure is closely related to the topology of quantum circuits, our framework also provides a basis for future studies on implementing parts of the tensor contraction procedure on quantum hardware.

The remainder of this paper is organized as follows. Section~\ref{sec:methodology} details the mathematical foundations of the MPS-VQC framework and the proposed optimization algorithms. Section~\ref{sec:experiments} describes the experimental setup and reports comparative results in terms of compression efficiency, training stability, and classification performance. Finally, Section~\ref{sec:conclusion} concludes the paper and discusses future directions.

\begin{figure}[t]
    \centering
    \includegraphics[width=\linewidth]{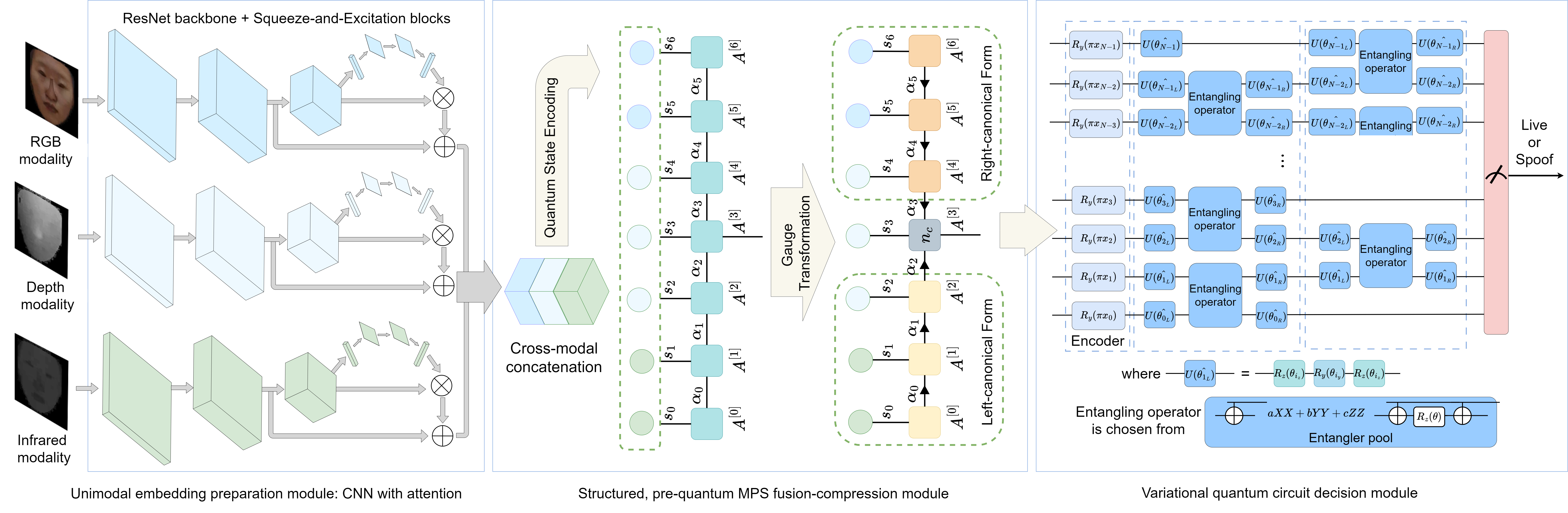}
    \caption{\textbf{Overview of the proposed hybrid quantum--classical multimodal face anti-spoofing framework.}
    RGB, Depth, and IR inputs are processed by modality-specific feature extractors with a Residual Network backbone and Squeeze-and-Excitation blocks to prepare unimodal embeddings. The embeddings are concatenated and fed into an MPS-based fusion and compression module, which performs structured fusion and differentiable dimensionality reduction and outputs a compact representation for quantum state encoding. A variational quantum circuit then performs the final classification between bona fide (live) and attack (spoof) samples.}
    \label{fig:framework}
\end{figure}

\section{Methodology}
\label{sec:methodology}
This section details the proposed hybrid quantum--classical framework for multimodal face anti-spoofing. As shown in Fig.~\ref{fig:framework}, the framework consists of three key components. (1) The unimodal embedding preparation module uses convolutional neural network encoders with a ResNet backbone and Squeeze-and-Excitation (SE) blocks to extract high-level unimodal embeddings from RGB, Depth, and IR inputs. (2) The structured, pre-quantum MPS fusion-compression module performs multimodal embeddings fusion and differentiable dimensionality reduction, producing a compact representation that fits the qubit budget of the variational quantum circuit. (3) The variational quantum circuit serves as the quantum classifier for final prediction. The remainder of this section describes the design and implementation details of these components.

\subsection{Unimodal Embedding Preparation Module}
This section outlines the classical data preprocessing and unimodal embedding construction process for multimodal face anti-spoofing. The objective of this stage is twofold: first, to maximally preserve fine-grained physical clues related to liveness; second, to suppress interference from nuisance factors such as pose, scale, misalignment, and environmental variations.

\noindent\textbf{Patch construction.} 
Given the synchronized multimodal inputs $\{I^{rgb}, I^{depth}, I^{ir}\}$, we construct patch-level representations to capture fine-grained spoofing cues. Compared with using full-face images, local patches are more effective at highlighting subtle spoofing artifacts. Specifically, from each modality we randomly sample $n$ fixed-size local patches from the face region to form the patch-level inputs. Each patch is represented as $\mathbf{p}^m_i \in \mathbb{R}^{3 \times h \times w}$, where $1 \leq i \leq n$ and $m \in \{rgb, depth, ir\}$.

\noindent\textbf{Unimodal feature encoding and dimension mapping.} 
To map high-dimensional multi-channel classical data into a low-dimensional feature space suitable for quantum processing, we construct independent convolutional neural network encoders $f_m(\cdot)$ for each modality. The backbone network adopts the ResNet architecture embedded with Squeeze-and-Excitation modules. Through the channel-wise attention recalibration mechanism, this structure enhances model robustness against partial facial occlusions such as masks and modality-specific noise. A projection head $g_m(\cdot)$, consisting of global pooling, linear layer, and a $\mathrm{tanh}$ activation function, is introduced at the end of the encoder to bridge the classical feature space and the quantum Hilbert space. The final unimodal embedding is obtained by:
\begin{equation}
\mathbf{e}^m = g_m\!\left(f_m(\mathbf{p}^m)\right) \in \mathbb{R}^{d}, \quad m \in \{rgb, depth, ir\}.\label{embedding}
\end{equation}
With these steps, we obtain independent embeddings for the three modalities.

\subsection{Structured, Pre-quantum MPS Fusion-Compression Module}
\label{subsec:mps-fusion}

To address the mismatch between high-dimensional embeddings and limited qubit resources of NISQ devices, we construct a differentiable quantum interface based on MPS. Serving as a bridge between the classical features and the quantum classifier, the MPS performs structured fusion of RGB, Depth, and IR features within the tensor network space. And then these features are mapped into a compact subspace that fits the budget of the subsequent variational quantum circuit.

\subsubsection{Quantum State Encoding}
Leveraging the capability of MPS to capture long-range correlations in one-dimensional sequences, we employ a cross-modal concatenation strategy to map the classical feature space onto a quantum spin chain. Specifically, we concatenate the unimodal embedding vectors $\mathbf{e}^m$ in Eq. (\ref{embedding}) to construct a global multimodal feature sequence $\mathbf{V}$:
\begin{equation}
\mathbf{V} = [\mathbf{e}^{rgb} \oplus \mathbf{e}^{depth} \oplus \mathbf{e}^{ir}] = [v_0, v_1, \dots, v_{L-1}],
\end{equation}
where $\oplus$ denotes vector concatenation, and the total sequence length $L$ equals the number of physical sites in the MPS. 

Adopting angle encoding, we construct the global tensor product state $|\Phi\rangle = \bigotimes_{n=1}^{L} |\phi^{[n]}\rangle$ as the input for the MPS network. Specifically, the local state $|\phi^{[n]}\rangle$ at each physical site is generated by applying a $R_Y$ rotation encoded with the corresponding feature value $v_n$:
\begin{equation}
|\phi^{[n]}\rangle = R_Y(2\pi v_n)|0\rangle = \cos(\pi v_n)|0\rangle + \sin(\pi v_n)|1\rangle. \label{quantumstate}
\end{equation}

\subsubsection{MPS Projector and Contraction}
\label{subsubsec:mps_projector}

We construct an MPS to project the high-dimensional input $|\Phi\rangle$ into a compact feature space. Mathematically, $\Psi$ decomposes a high-order global tensor into a contraction of $L$ low-order local tensors $\{A^{[n]}\}$ with an explicit feature output index $l$:
\begin{equation}
\Psi = \sum_{\alpha_0,\alpha_1,...,\alpha_{L}}  A^{[1]}_{\alpha_0 s_1 \alpha_1} \cdots A^{[n_c]}_{\alpha_{n_c-1} s_{n_c} \alpha_{n_c} l} \cdots A^{[L]}_{\alpha_{L-1} s_L \alpha_L}  \bigotimes_{n=0}^{L-1} | s_n \rangle,
\end{equation}
where $\{s_n\}$ denotes the physical indices of dimension $d$, corresponding to the local orthonormal basis. $\{\alpha_k\}$ denotes the virtual bond indices, whose bond dimension $\chi$ determines the model's expressive capacity and the limit of entanglement entropy. The index $l$, attached to the central tensor $A^{[n_c]}$, serves as the feature output index with a dimension set to $D_{fused}$.
This chain structure ensures that during contraction, the model not only processes intra-modal local structures but also utilizes virtual bonds as information channels to transfer and integrate context across different modal segments. This mechanism effectively captures deep semantic consistency between modalities.

By performing a full contraction between the physical indices of the input state $|\Phi\rangle$ and the MPS $\Psi$, we obtain the compressed feature $|\mathbf{h}\rangle$. This operation fuses the multimodal embeddings and projects them into a feature subspace spanned by the computational basis $\{|l\rangle\}$:
\begin{equation}
|\mathbf{h}\rangle =  \sum_{l} \left( \sum_{s} \Psi_{l, {s}} \Phi_{{s}} \right) |l\rangle.  \label{eq:operator_action}
\end{equation}
This dimensionality reduction effectively bridges the gap between high-dimensional classical data and the limited qubit capacity of the VQC.

However, for long-chain systems, performing direct global contraction with Eq. (\ref{eq:operator_action}) suffers from extremely high computational complexity and numerical instability. To address this, we adopt an optimized contraction strategy based on the mixed canonical form in our practical implementation, as detailed in the following subsection.

\begin{figure}[thbp]
    \centering
    \includegraphics[width=0.8\linewidth]{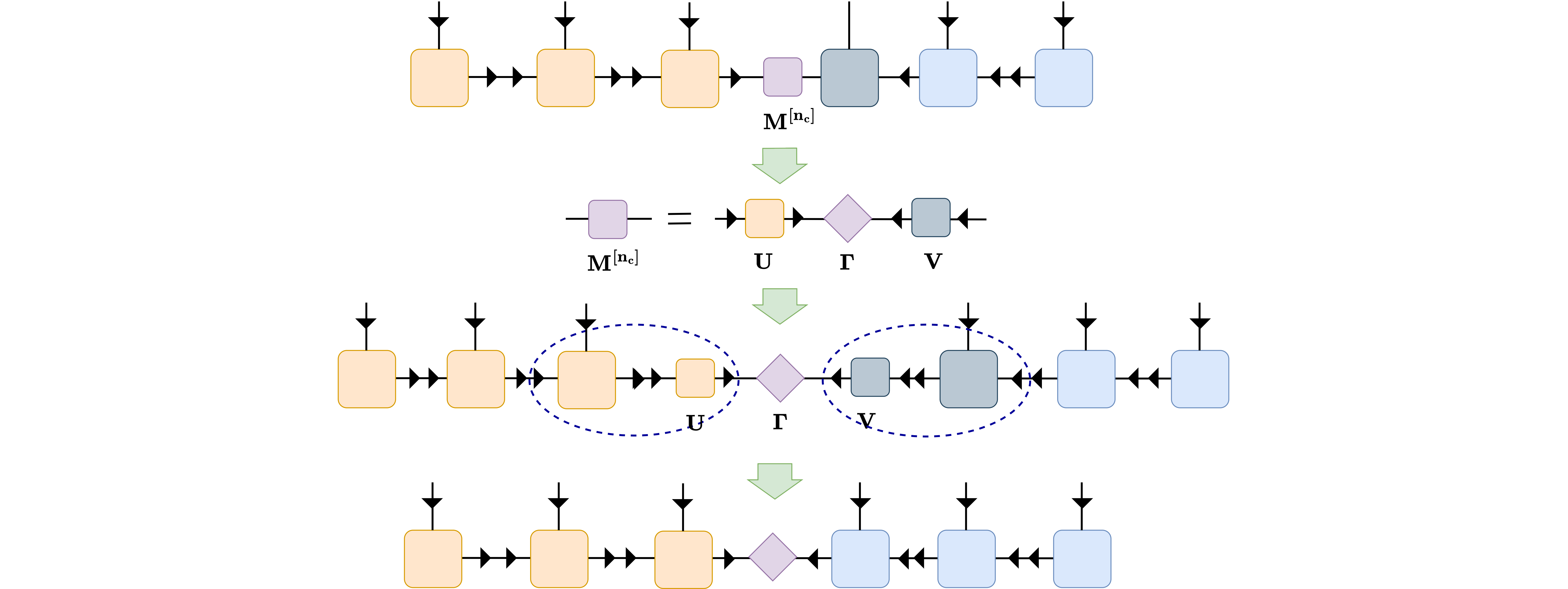}
    \caption{\textbf{Local SVD update for shifting the orthogonality center in a mixed-canonical MPS.} At site $n_c$, the center tensor is reshaped into a matrix $M^{[n_c]}$ and factorized as $M^{[n_c]}=U\,\Gamma\,V$. After truncating the singular spectrum to a target virtual bond dimension, $U$ is absorbed into the left-canonical part, while the product $V$ is contracted into the adjacent right tensor. This operation transfers the orthogonality center from site $n_c$ to $n_c+1$ and updates the virtual bond dimension.}
    \label{fig:mps_svd}
\end{figure}

\subsubsection{Mixed Canonical Form and Virtual Dimension Truncation}
\label{subsubsec:mps_pruning}

To address the numerical instability and computational redundancy associated with direct global tensor contraction in long chains, we adopt a linear scan update strategy based on the mixed canonical form. Instead of processing the entire network simultaneously, this strategy maintains a moving orthogonality center, transforming global computations into a sequence of local tensor updates. This mechanism ensures that the local tensors of the MPS satisfy orthonormal constraints throughout training and inference. This not only mitigates numerical drift but also provides a clear truncation window, allowing us to control the model's entanglement capacity and computational cost by regulating the virtual bond dimension.

\noindent\textbf{Mixed Canonical Constraints.}
Without canonical constraints, MPS models suffer from gauge ambiguity, where arbitrary perturbations in local tensors can lead to non-linear drifts in the global wavefunction norm. Consequently, gradient calculations cannot decouple the current site from the background environment, necessitating the processing of full-chain correlations, which severely undermines numerical stability.
To mitigate this issue, we strictly maintain the MPS in the mixed canonical form, characterized by a unique orthogonality center at site $n_c$. The orthogonality center partitions the network into two segments: tensors to the left of $n_c$ satisfy the left-canonical condition $A^\dagger A = I$, while those to the right adhere to the right-canonical condition $A A^\dagger = I$. In this structure, the orthogonality center contains the singular spectrum and global norm information of the entire state, whereas the remaining tensors act as isometries that construct orthonormal bases for the local feature space. This property allows us to safely decompose global optimization into local variational updates around the orthogonality center.

\noindent\textbf{Local Tensor Update and Virtual Dimension Truncation.}
Based on the mixed canonical form, we employ a linear scan update strategy to optimize the MPS parameters. As illustrated in Fig. \ref{fig:mps_svd}, the core of this strategy utilizes Singular Value Decomposition (SVD) to achieve the site-by-site migration of the orthogonality center and dynamic adjustment of dimensions. The SVD explicitly decomposes the local tensor into Schmidt modes, allowing us to implement a truncation of the virtual bond dimension while maintaining the optimal approximation of the model. We exemplify this process by describing the shift of the orthogonality center from site $n_c$ to $n_c+1$:
\begin{itemize}
    \item Step 1: Tensor Reshaping and SVD. At the orthogonality center, we reshape the local tensor at site $n_c$ into a matrix $M^{[n_c]}_{(\alpha_{n_c-1} s_{n_c}),\,\alpha_{n_c}}$ by merging the left virtual index $\alpha_{n_c-1}$ and the physical index $s_{n_c}$ into a composite row index and using the right virtual index $\alpha_{n_c}$ as the column index. We then compute the SVD
    \begin{equation}
        M^{[n_c]}_{(\alpha_{n_c-1} s_{n_c}),\,\alpha_{n_c}}
        = \sum_{\gamma} U_{(\alpha_{n_c-1} s_{n_c}),\,\gamma}\,
        \Gamma_{\gamma}\,
        V_{\gamma,\,\alpha_{n_c}},
    \end{equation}
    where $U$ is left-isometric, $V$ is right-isometric, and $\Gamma$ denotes the non-negative singular spectrum in descending order. The squared singular values $\Gamma_{\gamma}^2$ correspond to the Schmidt coefficients at this bipartition and equal the eigenvalues of the reduced density matrix.

    \item Step 2: Virtual Bond Dimension Truncation. To regulate computational complexity, we establish a global maximum virtual bond dimension denoted as $\chi_{\text{set}}$. According to the Eckart-Young-Mirsky theorem, retaining the top $\chi_{\text{set}}$ largest singular values yields the optimal rank-$\chi_{\text{set}}$ approximation of the original tensor in terms of the Frobenius norm. Specifically, we preserve the leading $\chi' = \min(D_{\text{actual}}, \chi_{\text{set}})$ singular values and their corresponding singular vectors. This operation eliminates singular value components with negligible contributions and compels the model to approximate the original tensor using only the most significant $\chi'$ feature vectors.

    \item Step 3: Absorption and Center Shift. After truncation, the factor $U_{:,1:\chi'}$ is reshaped back and absorbed into site $n_c$, so that the tensor at site $n_c$ becomes left-canonical. The product $(\Gamma V)_{1:\chi',:}$, which carries the singular spectrum and norm information, is then contracted into the adjacent right tensor:
    \begin{equation}
        \tilde{A}^{[n_c+1]}_{\gamma s_{n_c+1} \alpha_{n_c+1}}
        = \sum_{\alpha_{n_c}}
        (\Gamma V)_{\gamma \alpha_{n_c}}\,
        A^{[n_c+1]}_{\alpha_{n_c} s_{n_c+1} \alpha_{n_c+1}},
    \end{equation}
    where $\gamma\in\{1,\dots,\chi'\}$ becomes the updated virtual bond index. This contraction shifts the orthogonality center from $n_c$ to $n_c+1$ and completes one local update step.
\end{itemize}

\subsubsection{Nonlinear Extension of MPS}
\label{subsubsec:activated_mps}
Although standard MPS provides a rigorous framework for sequence compression, relying solely on linear tensor contractions often limits expressivity in deep learning contexts. Specifically, linear structures struggle to model complex non-linear decision boundaries and face challenges in propagating gradients effectively across long sequences. Consequently, we introduce a non-linear extension, denoted as Activated MPS, as a robust alternative.

The vector $v_L^{[n-1]}$ represents the contraction result of the tensor network from the first site up to site $n-1$, corresponding to the virtual bond index $\alpha_{n-1}$.
For the current site $n$, the input is the local quantum state $|\phi^{[n]}\rangle$ defined in Eq.~(\ref{quantumstate}). We denote its coefficient in the computational basis $|s_n\rangle$ as $\phi^{[n]}_{s_n} = \langle s_n | \phi^{[n]} \rangle$. The update rule for the forward propagation is formulated as:
\begin{equation}
\Delta v_{\alpha_n} = \sum_{\alpha_{n-1}, s_n} A^{[n]}_{\alpha_{n-1} s_n \alpha_n} v^{[n-1]}_{\alpha_{n-1}} \phi^{[n]}_{s_n} + b^{[n]}_{\alpha_n},
\label{eq:activated_contraction}
\end{equation}
\begin{equation}
v^{[n]}_{\alpha_n} = v^{[n-1]}_{\alpha_n} + \sigma_{\mathrm{t}}(\Delta v_{\alpha_n}),
\label{eq:activated_residual}
\end{equation}
where $A^{[n]}_{\alpha_{n-1} s_n \alpha_n}$ is the trainable local tensor consistent with the notation in Eq.~(\ref{eq:operator_action}), and $b^{[n]}_{\alpha_n}$ is a learnable bias. Here, $\sigma_{\mathrm{t}}(\cdot)$ denotes the $\mathrm{tanh}$ activation function, distinguished from the readout activation used later.
As shown in Eq.~(\ref{eq:activated_residual}), the residual addition implies a direct identity mapping between the vector spaces of $\alpha_{n-1}$ and $\alpha_n$, allowing gradients to flow unimpeded along the virtual bond dimension. The activation introduces non-linearity, enabling the model to capture complex, non-linearly separable patterns.

\subsubsection{Numerical Stability and Dynamic Normalization}

When the MPS chain length $L$ is large, successive matrix multiplications can cause the norms of intermediate tensors to decay or explode exponentially with $L$, leading to floating-point underflow or vanishing gradients. To address this, we introduce a dynamic normalization mechanism with $\varepsilon$-smoothing during sweeping and contraction.
For any intermediate contraction tensor or local tensor, denoted generally as $T$, we perform the following normalization:
\begin{equation}
T \leftarrow \frac{T}{\|T\|_F + \varepsilon},
\end{equation}
where $\varepsilon \approx 10^{-6}$ is a numerical stability term.
This strategy ensures that the norms of all intermediate tensors remain around $1.0$. Unlike simple normalization, the inclusion of $\varepsilon$ prevents division-by-zero errors caused by zero tensors and ensures that the tensor magnitude exhibits linear rather than exponential gradient behavior during backpropagation. This fundamentally mitigates numerical instability in training deep tensor networks and guarantees the effective propagation of cross-modal features along long chains.

\subsection{Variational Quantum Circuit Decision Module}
\label{subsec:vqc-circuit}

To preserve the one-dimensional entanglement structure of the preceding MPS, we construct a variational quantum circuit based on the Unitary Tensor Train (UTT) topology. This architecture physically instantiates the abstract tensor network, ensuring a unified topology throughout the classifier. The circuit comprises two functional stages: feature encoding and variational evolution.

To bridge the MPS output with the quantum Hilbert space, the fused and compressed feature $\mathbf{h}$ undergoes dimensional adaptation via a classical fully connected layer. Its components are then mapped to rotation angles of $R_Y$ gates acting on the initial state. The resulting encoded state serves as the input for the subsequent evolution:
\begin{equation}
    |\psi_{\mathrm{enc}}\rangle = \bigotimes_{j=1}^{N_q} R_Y(\theta_j(\mathbf{h})) |0\rangle_j,  \label{eq:encoding}
\end{equation}
where $\boldsymbol{\theta}(\mathbf{h}) = \mathbf{W}\mathbf{h} + \mathbf{b}$.

Subsequently, the variational ansatz is structurally defined as an ordered cascade of $N_q$ local unitary transformations, comprising $N_q-1$ two-qubit entanglement blocks followed by a terminal single-qubit gate:
\begin{equation}
\label{eq:ansatz_evolution}
|\psi\rangle = \mathcal{U}^{[N_q-1]} \mathcal{U}^{[N_q-2]} \cdots \mathcal{U}^{[0]} |\psi_{\mathrm{enc}}\rangle.
\end{equation}
 For $0 \le n \le N_q-2$, the two-qubit blocks $\mathcal{U}^{[n]}$ are constructed according to the Cartan Decomposition. Mathematically, each block is factorized as:
\begin{equation}
\mathcal{U}^{[n]} = \left( U(\boldsymbol{\theta}_{n_L}) \otimes U(\boldsymbol{\theta}_{n+1_L}) \right)  \mathcal{O}^{[n]}_{e}  \left( U(\boldsymbol{\theta}_{n_R}) \otimes U(\boldsymbol{\theta}_{n+1_R}) \right),  \label{eq:cartan_decomposition}
\end{equation}
where $\mathcal{O}^{[n]}_{e}$ represents an entanglement operator selected from a predefined pool $\mathcal{P}_e = \{ \text{CNOT}, \text{CRZ}, \alpha_x XX + \alpha_y YY + \alpha_z ZZ \}$. This set encompasses both fixed logic gates and parameterized Heisenberg interactions, offering diverse qubit coupling modes. For both the terminal single-qubit transformation $\mathcal{U}^{[N_q-1]}$ and the local operations $U(\boldsymbol{\phi})$ in Eq. (\ref{eq:cartan_decomposition}), we uniformly adopt the universal decomposition form:
\begin{equation}
U(\boldsymbol{\theta}) = e^{i\theta_{p}} R_Z(\theta_{z1}) R_Y(\theta_{y}) R_Z(\theta_{z2}).
\end{equation}
This $Z$-$Y$-$Z$ rotation sequence allows for the traversal of the entire Bloch sphere, achieving arbitrary local basis transformations. Finally, upon completion of the circuit evolution, we perform Pauli-Z measurements on specific qubits. The expectation values are processed through a fully connected layer with a $\operatorname{sigmoid}$ activation to yield the final classification probability.

\begin{figure}[ht]
    \centering
    \includegraphics[width=\textwidth]{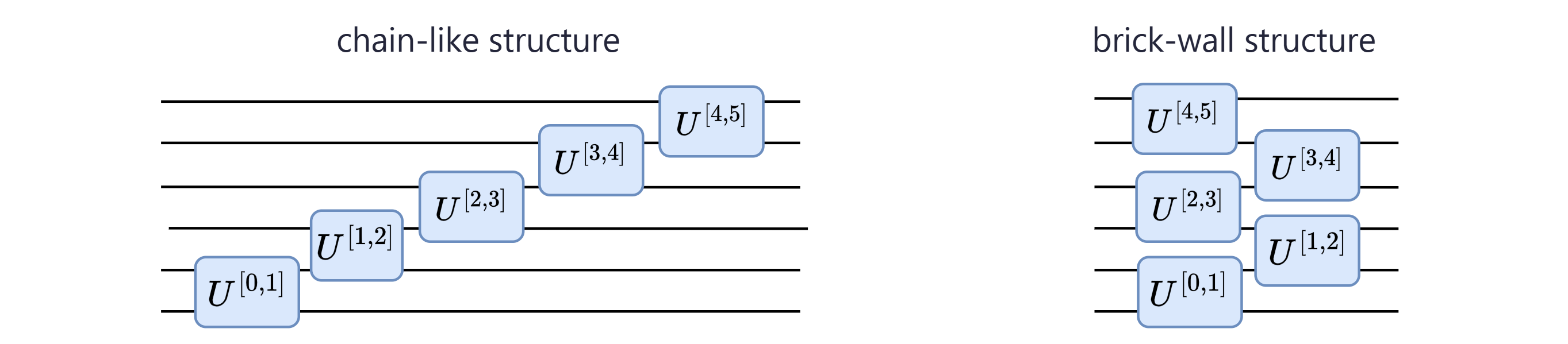} 
    \caption{\textbf{Chain-like and brick-wall structure circuit diagrams.} The blocks $U^{[i,j]}$ represent two-qubit gates acting on qubits $i$ and $j$.}
    \label{fig:circuit_topologies}
\end{figure}

It is worth noting that the UTT architecture possesses inherent flexibility, allowing it to support various connectivity patterns, such as the brick-wall structure, as illustrated in Fig.~\ref{fig:circuit_topologies}. While the chain-like structure aligns well with the MPS representation, the brick-wall structure is often favored for its potential in achieving faster entanglement propagation in quantum hardware.

The subsequent error analysis confirms the theoretical robustness of our approach, demonstrating the approximate equivalence between these two structures. The dominant source of error in sequentially applied unitary operations, particularly when approximating continuous time evolution, typically arises from the Trotter-Suzuki (TS) decomposition. This introduces errors of \( O(\tau^2) \) due to the non-commutativity of local operations. When quantitatively comparing the chain-like and brick-wall structures, considering two local evolution operators \( A \) and \( B \) and a conceptual training step size \(\tau\), the error difference between them can be expressed as:
\[
    \begin{aligned}
    e^{\tau (A+B)}-e^{\tau A}e^{\tau B} &= \tau^2 (\frac{1}{2}AB +\frac{1}{2}BA-AB)+O(\tau^3) \\
    &\approx -\frac{\tau^2}{2}[A,B] 
    =O(\tau^2),
    \end{aligned}
\]
where \([A,B]\) denotes the commutator of \(A\) and \(B\). This analysis indicates that the structural discrepancy between chain-like and brick-wall arrangements also introduces an error of \( O(\tau^2) \), which is consistent with the leading-order TS error. Consequently, these structures can be considered approximately equivalent in terms of their theoretical error contributions.

\subsection{Two-Stage Decoupled Optimization}
\label{subsec:training-strategy}

While end-to-end joint training theoretically offers the potential for global optimality, hybrid quantum-classical architectures encounter stability challenges in practice. The sequential contraction structure of the MPS creates high sensitivity to numerical fluctuations in input data. Specifically, when processing multimodal data with outliers or high variance, a long MPS chain can induce exponential gradient explosion or vanishing. Furthermore, large adjustments in MPS parameters during joint updates cause non-stationary input distributions for the subsequent quantum circuit. This continuous drift in the feature space prevents the VQC from maintaining a stable gradient descent direction, often resulting in oscillation or divergence of the optimization trajectory. To address these, we adopt a two-stage decoupled optimization strategy. This approach temporally separates the feature extraction of the MPS from the quantum evolution of the VQC, stabilizing the input-output behavior of each module sequentially.

The first stage focuses on training the MPS fusion layer to map multimodal inputs into low-dimensional features $\mathbf{h}$ shown in Eq. (\ref{eq:operator_action}). In this stage, the quantum circuit is bypassed, and the MPS output connects directly to a temporary classical fully connected layer $\mathbf{W} \in \mathbb{R}^{K \times D_{\text{fused}}}$, where $K$ is the number of classes. We minimize the cross-entropy loss to extract feature representations with high class discriminability. For a given sample, the loss function is:
\begin{equation}
\label{eq:loss_phase1}
\mathcal{L}_1 = -\sum_{k=1}^{K} y_{k} \log P(k|\mathbf{h}),
\end{equation}
where $\{y_k\}$ represents the one-hot encoding of the ground truth label. The prediction probability $P(k|\mathbf{h})$ is computed by the Softmax function:
\begin{equation}
    P(k|\mathbf{h}) = \frac{e^{(\mathbf{W} \mathbf{h})_k}}{\sum_{j=1}^{K} e^{(\mathbf{W} \mathbf{h})_j}}.
\end{equation}
Minimizing $\mathcal{L}_1$ enhances the feature response for the correct class and suppresses the logits of non-target classes, thereby improving the distinguishability of the MPS output features.

To mitigate gradient instability caused by tensor network contraction, we apply gradient clipping. During backpropagation, if the $L_2$ norm of the gradient vector $\mathbf{g}$ exceeds a threshold $\delta$, it is renormalized:
\begin{equation}
\mathbf{g} \leftarrow \begin{cases} 
\delta \cdot \frac{\mathbf{g}}{\|\mathbf{g}\|_2}, & \text{if } \|\mathbf{g}\|_2 > \delta \\
\mathbf{g}, & \text{otherwise}
\end{cases}.
\end{equation}
This method limits gradient anomalies from data noise, ensuring the MPS maps high-dimensional inputs to compact, linearly separable feature vectors.

In the second stage, all MPS parameters are frozen and serve as a fixed feature extractor that maps multimodal embeddings to static feature vectors $\mathbf{h} \in \mathbb{C}^{D_{\text{fused}}}$. These features are encoded into quantum states $|\psi_{\mathrm{enc}}\rangle$ using the protocol in Eq. (\ref{eq:encoding}). The optimization goal is to identify the measurement basis in the Hilbert space that maximally distinguishes the quantum state distributions of ``live'' and ``spoof'' classes.

Following the quantum evolution described in Eq. (\ref{eq:ansatz_evolution}), we perform Pauli-Z measurements on all readout qubits to obtain the expectation vector $\mathbf{E} = [\langle Z_0 \rangle, \dots, \langle Z_{N_q-1} \rangle]^T$. This vector is processed by trainable weights $\mathbf{w}$ and a bias $b$, then mapped to a prediction probability via a $\operatorname{sigmoid}$ function $\sigma(\cdot)$:
\begin{equation}
P(y=1|\mathbf{h}) = \sigma\left( \sum_{j=0}^{N_q-1} w_j \langle \psi | Z_j | \psi \rangle + b \right),
\end{equation}
where $w_j$ is the weight for the $j$-th qubit. This weighted readout strategy enables the VQC to select qubit channels with high discriminative information. The optimization objective $\mathcal{L}_{2}$ uses the same cross-entropy form as Eq. (\ref{eq:loss_phase1}) but updates only the VQC variational parameters $\boldsymbol{\theta}$ and classical readout parameters. With the stable feature manifold from Stage I, the quantum circuit avoids input distribution drift and converges to the decision boundary.

To further improve convergence quality, we incorporate a cosine annealing with warm restarts strategy in both stages. This mechanism adjusts the learning rate $\eta_t$ based on the current step $t$:
\begin{equation}
\eta_t = \eta_{\min} + \frac{1}{2}(\eta_{\max} - \eta_{\min}) \left( 1 + \cos\left( \pi \frac{t}{T_r} \right) \right),
\end{equation}
where $\eta_{\max}$ and $\eta_{\min}$ are the maximum and minimum learning rates, and $T_r$ is the total steps in the current restart cycle. Periodically resetting the learning rate assists the model in escaping local saddle points and supports the optimization of both modules.

\section{Experimental Results and Analysis}
\label{sec:experiments}

We conducted experiments on the standard dataset CASIA-SURF \cite{zhang2019dataset} to validate the effectiveness of the hybrid MPS-VQC framework in multimodal face anti-spoofing task. This section provides an analysis of the physical principles and engineering characteristics of the coupled system, focusing on four core dimensions: feature fusion compression efficiency, model parameter complexity, optimization dynamics stability, and low-dimensional quantum discriminative capability.

\subsection{Experimental Setup}
\label{subsec:setup}

The experiments in this work are implemented using a hybrid classical-quantum architecture. The classical components, comprising backbones for unimodal feature extraction and the MPS tensor network for multimodal fusion and compression, are built upon the PyTorch~\cite{paszke2019pytorch} framework. To address the high computational demands of tensor contraction, these components are accelerated using the CUDA 12.4 environment on the computing platform equipped with NVIDIA Tesla V100S GPUs. All numerical simulations of the quantum circuits are implemented using the MindSpore Quantum framework~\cite{mindspore_quantum}.

In the data preprocessing stage, we adopt a patch-based sampling strategy, extracting local regions of size $32 \times 32$ from the original RGB, Depth, and IR images to construct the input dataset. All modalities are transferred into feature embeddings of dimension $D_{emb}$ via parameter-frozen ResNet and SE-Net backbones. The physical chain length of the MPS is set to $L_{MPS} = 3 \times D_{emb}$ to accommodate the cascaded tri-modal input. Detailed structures of the variational quantum circuit are described in Section \ref{subsec:vqc-circuit}.

Model performance evaluation follows the ISO/IEC 30107-3 standard protocol. We report three core metrics: Attack Presentation Classification Error Rate (APCER), which is the proportion of attack samples misclassified as bona fide; Bona Fide Presentation Classification Error Rate (BPCER), the proportion of bona fide samples misclassified as attacks; and the Average Classification Error Rate (ACER). ACER reflects the balance between defense capability and user experience, and it is defined as:
\begin{equation}
    ACER = \frac{APCER + BPCER}{2}.
\end{equation}
Additionally, to assess performance in high-security scenarios, we report the True Positive Rate (TPR) at a fixed False Positive Rate (FPR) of $10^{-1}\%$, denoted as \textbf{TPR@FPR=$10^{-1}$}. Lower ACER and higher TPR values indicate superior defense performance.

\subsection{Evaluation of Performance Benchmarks and Parameter Efficiency}
\label{subsec:performance_complexity}

This section aims to validate the capability of the proposed MPS-VQC architecture as an efficient interface between high-dimensional classical features and NISQ devices, focusing on discrimination accuracy, physical mechanisms, and computational resource consumption. Considering the limited number of logical qubits in NISQ devices, we focus on the comprehensive performance when compressing high-dimensional multimodal features into an extremely low dimension of $D_{fused}=4$. We compare the MPS-VQC framework against powerful classical  neural network baselines, including ResNet, VGG16, GoogleNet, and AlexNet. Table \ref{tab:model_comparison} presents the discrimination accuracy, while Table \ref{tab:complexity_comparison} reveals the model complexity required to achieve such performance.

\begin{table*}[tbp]
\centering
\caption{Performance comparison of different models on the CASIA-SURF dataset.}
\label{tab:model_comparison}

\begin{threeparttable}
\resizebox{\textwidth}{!}{%
\begin{tabular}{@{} l cccccc @{}} 
\toprule
Model Name & \multicolumn{2}{c}{\textbf{3×256 $\to$ 4 Dim (\%)}} & \multicolumn{2}{c}{\textbf{3×128 $\to$ 4 Dim (\%)}} & \multicolumn{2}{c}{\textbf{3×64 $\to$ 4 Dim (\%)}} \\
\cmidrule(lr){2-3} \cmidrule(lr){4-5} \cmidrule(lr){6-7}
& ACER & TPR@FPR=0.1 & ACER & TPR@FPR=0.1 & ACER & TPR@FPR=0.1 \\
\midrule
ResNet~\cite{he2016deep} & 0.3818 & 100.00 & 0.5141 & 100.00 & 0.4650 & 100.00 \\
AlexNet~\cite{krizhevsky2012imagenet} & 0.2759 & 100.00 & 0.4800 & 99.97 & 0.5594 & 99.98 \\
GoogleNet~\cite{szegedy2015going} & \textbf{0.1701} & 100.00 & \textbf{0.1663} & 100.00 & 0.4423 & 100.00 \\
VGG16~\cite{simonyan2015very} & 0.6388 & 100.00 & 0.3175 & 100.00 & \textbf{0.3138} & 100.00 \\
\textbf{Activated MPS}\tnote{1} & 0.7869 & 92.95 & 0.4385 & 99.97 & 0.6879 & 99.06 \\
\textbf{Standard MPS}\tnote{1}  & 0.6517 & 99.90 & 0.6350 & 100.00 & 1.7387 & 99.73 \\
\textbf{Activated MPS}\tnote{2} & \textbf{0.2117} & 100.00 & 0.4309 & 100.00 & \textbf{0.6501} & 99.93 \\
\textbf{Standard MPS}\tnote{2}  & 0.3115 & 99.93 & \textbf{0.3251} & 100.00 & 1.3985 & 99.87 \\
\bottomrule
\end{tabular}%
}
\begin{tablenotes}[flushleft]
\footnotesize
\item[1] MPS trained without virtual dimension truncation.
\item[2] MPS trained with virtual dimension truncation.
\end{tablenotes}
\end{threeparttable}
\end{table*}

\noindent\textbf{Dimensionality Dependence and Structured Fusion Advantage.} 
Experiments reveal a significant trend: as the input embedding dimension decreases from $3 \times 256$ to $3 \times 64$, the advantage of MPS over classical neural network baselines gradually diminishes. In the high-dimensional regime of $3 \times 256$, MPS demonstrates superior structured fusion capabilities. Facing long sequence features, classical neural network baselines are often limited by local receptive fields as seen in ResNet or parameter redundancy such as that found in AlexNet, making it difficult to efficiently capture cross-modal long-range correlations. In contrast, MPS constructs a global information transmission channel via the virtual bond dimension, enabling holistic modeling of high-dimensional sequences through the long-range entanglement mechanism with minimal parameters.
However, when the dimension drops to $3 \times 64$, the performance of MPS declines. We attribute this to the physical property of correlation length: in low-dimensional spaces, features have been highly compressed and localized by the backbone, truncating the original long-range correlations. Consequently, the data structure aligns better with the local inductive bias of the convolutional operations rather than the entanglement structure of MPS. This contrast conversely validates the unique advantage of the MPS-VQC framework in multimodal fusion tasks involving high-dimensional, long-range dependencies.

\noindent\textbf{High-dimensional Entanglement Capture with Extreme Parameter Efficiency.}
Under an extreme compression ratio of nearly 200:1—specifically reducing dimensions from $3 \times 256$ to 4—the Activated MPS with truncation strategy achieves an ACER of $0.2117\%$. This performance significantly outperforms the classical baselines, including ResNet at $0.3818\%$, VGG16 at $0.6388\%$, and AlexNet at $0.2759\%$. More importantly, the model demonstrates exceptional parameter efficiency. As shown in Table \ref{tab:complexity_comparison}, Activated MPS requires only $\sim$0.25M parameters to surpass AlexNet, which utilizes nearly 40M parameters. This stark contrast provides strong evidence that tensor networks possess a higher information expression density than classical neural networks based on local receptive fields or fully connected layers when processing high-dimensional long-range entanglement.

\begin{table*}[htbp]
    \centering
    \begin{threeparttable} 
        \caption{Parameter count comparison of different fusion models.}
        \label{tab:complexity_comparison}
        \begin{tabular}{@{}lccccc@{}}
            \toprule
            \textbf{Model} & AlexNet~\cite{krizhevsky2012imagenet} & VGG16~\cite{simonyan2015very} & ResNet~\cite{he2016deep} & GoogleNet~\cite{szegedy2015going} & \textbf{MPS} \\
            \midrule
            \textbf{Params} & $\sim$45.17 M & $\sim$11.81 M & $\sim$11.75 M & $\sim$0.49 M & \textbf{$\sim$0.25 M} \\
            \botrule
        \end{tabular}

        \begin{tablenotes}
            \footnotesize 
            \item Note: M denotes million. The MPS parameter count is estimated with truncated bond dimension $\chi=4$, a setting where it achieves performance comparable to the classical baselines.
        \end{tablenotes}
    \end{threeparttable}
\end{table*}
\noindent\textbf{Advantages of Truncation Strategy from a Renormalization Group Perspective.} 
Table \ref{tab:model_comparison} confirms the critical role of virtual bond dimension truncation. The strategy of high-dimensional initialization followed by SVD truncation significantly outperforms direct training in low bond dimension spaces. From the perspective of Renormalization Group (RG) flow, high-dimensional initialization allows the model to explore a larger entanglement entropy space initially, capturing complex long-range correlations; the subsequent singular value truncation acts similarly to coarse-graining in RG, effectively stripping away redundant high-frequency noise while retaining key features essential for the classification task. This confirms that virtual dimension truncation is a key technique for effectively mapping classical high-dimensional features to low-dimensional quantum states.

\begin{figure*}[tbp]
    \centering
    \begin{minipage}[b]{0.48\textwidth}
        \centering
        \includegraphics[width=\linewidth]{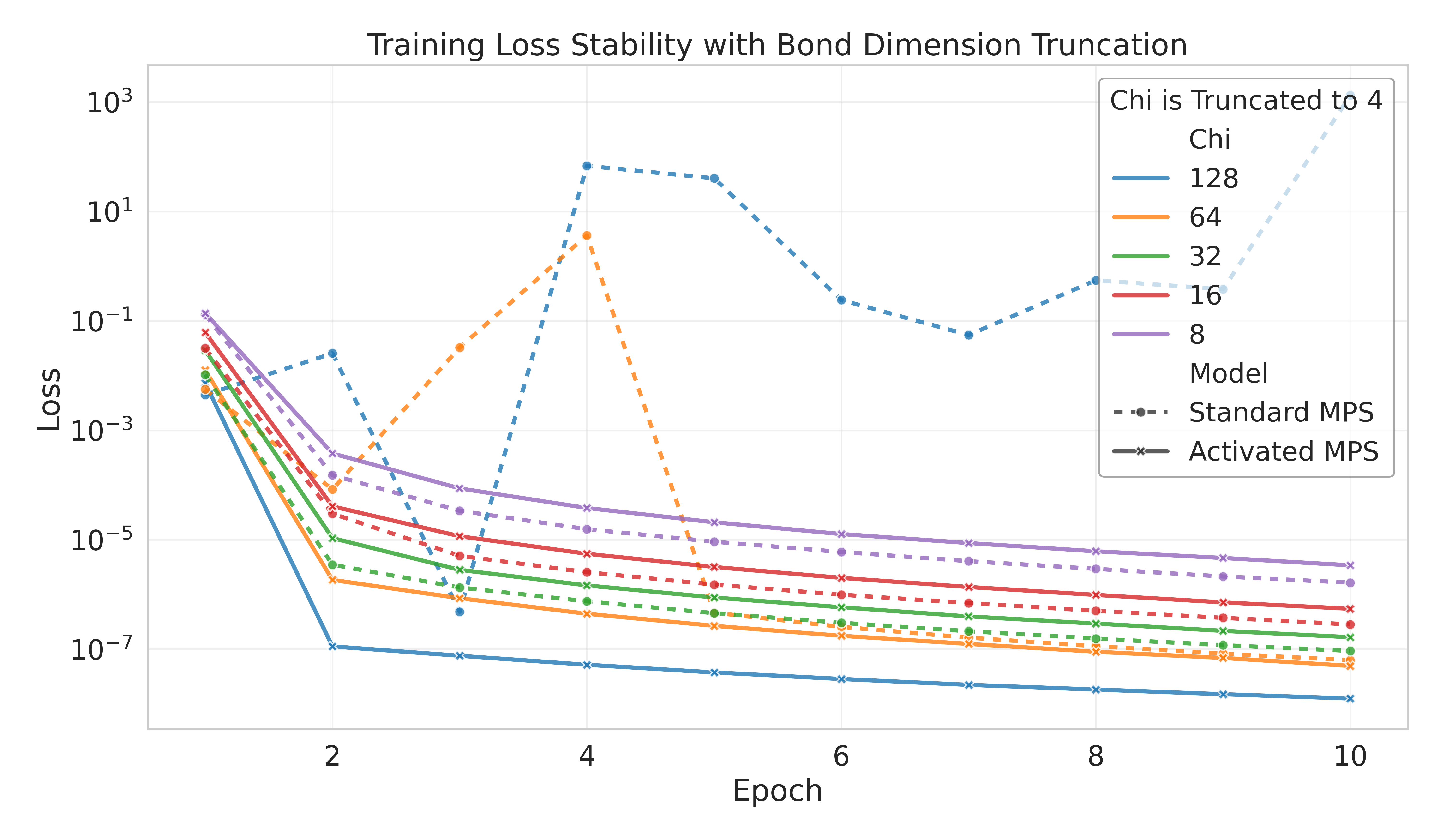}
        \centerline{(a)} 
    \end{minipage}
    \hfill
    \begin{minipage}[b]{0.48\textwidth}
        \centering
        \includegraphics[width=\linewidth]{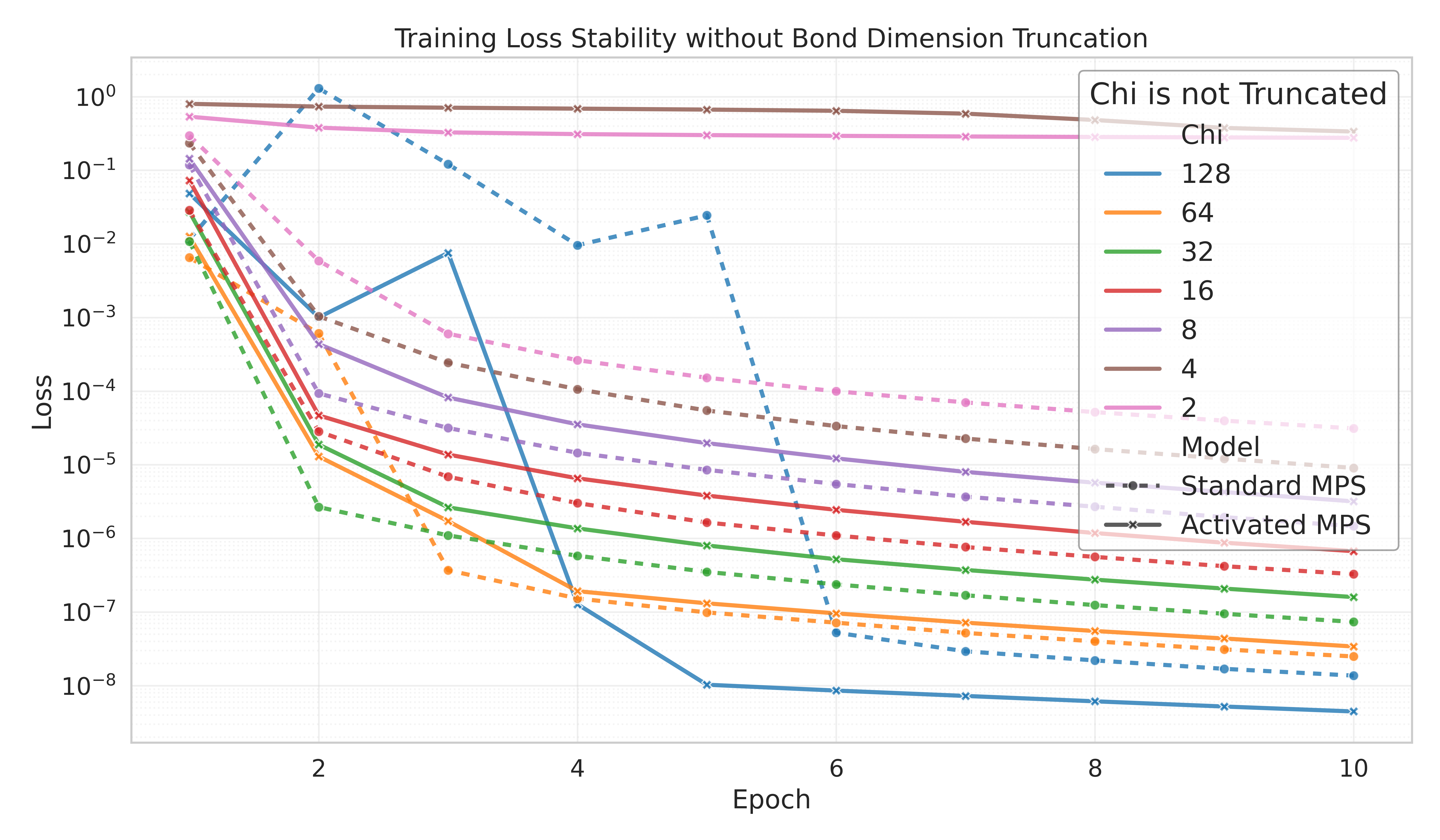}
        \centerline{(b)} 
    \end{minipage}
    
    \caption{\textbf{Training loss stability comparison.} \textbf{(a)} Training loss for models with nominal bond dimensions $\chi$ that are truncated during training to a maximum virtual bond dimension $\chi_{\max}=4$. \textbf{(b)} Training loss for the same nominal $\chi$ values without any bond dimension truncation. In both panels, dashed lines denote Standard MPS and solid lines denote Activated MPS; Activated MPS shows smoother convergence.}
    \label{fig:loss_comparison}
\end{figure*}

\noindent\textbf{Optimization Dynamics and Model Stability.}
Beyond static performance advantages, we further verified the dynamic stability of the training process via loss convergence analysis. As illustrated in Fig. \ref{fig:loss_comparison}(a) for the extreme compression scenario $D_{fused}=4$ and Fig. \ref{fig:loss_comparison}(b) for the unconstrained scenario characterized by pure long-range correlations, the Standard MPS initialized with high $\chi$ exhibits severe loss spikes with fluctuations reaching magnitudes of $10^0$. This indicates that Standard MPS, relying on linear tensor decomposition, struggles to project high-complexity manifolds into low-dimensional subspaces or propagate gradients along long tensor chains. In contrast, Activated MPS demonstrates superior robustness in both cases, with loss curves converging smoothly and fluctuations restricted to the $10^{-2}$ magnitude. This confirms that the nonlinear activation mechanism acts as a "gating" regulator within the tensor network, effectively suppressing gradient vanishing and exploding issues to ensure controllable convergence when processing complex high-dimensional distributions.

\noindent\textbf{Balance of Accuracy-Complexity-Quantum Compatibility.}
Although some classical neural network baselines, such as GoogleNet, perform excellently in specific metrics, they are essentially classical black-box models whose deep nonlinear structures are extremely difficult to map onto quantum hardware. In contrast, MPS possesses native quantum mapping capabilities, as its tensor contraction process naturally corresponds to unitary transformations and measurements in quantum circuits. Consequently, Activated MPS achieves a level of quantum readiness unattainable by classical neural networks while maintaining high accuracy. This balance of accuracy and quantum compatibility, achieved at an extremely low parameter cost, makes it an ideal bridge connecting classical data with NISQ devices.

\subsection{Analysis of Quantum Circuit Scale and Data Efficiency}
\label{subsec:quantum_robustness}

\begin{figure}[tbp]
    \centering
    \includegraphics[width=\linewidth]{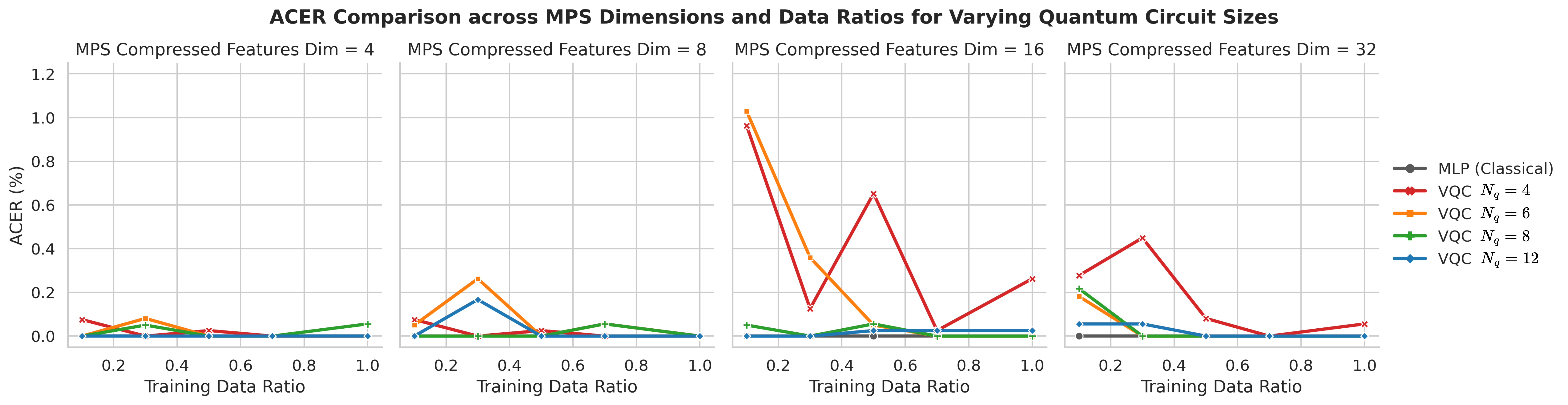}
    \caption{ACER performance comparison across different MPS-compressed feature dimensions and training data ratios. From left to right, the subplots show ACER evaluated for models trained with MPS-compressed features with dimension $D_{\mathrm{fused}}=4,8,16,32$, respectively. Each colored line corresponds to one model: the classical MLP baseline or a VQC with $N_q=4,6,8,12$ qubits. The training data ratio takes the values 0.1, 0.3, 0.5, 0.7, and 1.0, corresponding to using 10\%, 30\%, 50\%, 70\%, and 100\% of the training set, respectively.}
    \label{fig:data_efficiency}
\end{figure}

Following the verification of the feature compression capabilities of the MPS structure, we further evaluated the performance of the backend VQC classifier under varying quantum resource configurations and data scales. In this experiment, the original $3 \times 128$-dimensional multimodal embeddings were projected into feature vectors with dimensions $D_{fused} \in \{4, 8, 16, 32\}$ via MPS, which then served as inputs for the VQC. To investigate the impact of different model configurations, we set the number of qubits $N_q$ for the VQC to 4, 6, 8, and 12, and trained the models using random subsets ranging from 10\% to 100\% of the full training set. Fig. \ref{fig:data_efficiency} illustrates the ACER performance curves under these different experimental settings. The overall experimental results indicate that both the classical multilayer perceptron (MLP) and the various VQC configurations maintained low error rates across most testing conditions. The ACER values remained generally below $1.25\%$, suggesting that after the MPS compresses the original 384-dimensional features down to dimensions between 4 and 32, the output features retain the primary discriminative information necessary to distinguish between live and spoof samples. This reduction in feature space dimensionality allows the subsequent classifiers to be trained effectively with lower-dimensional inputs.

As shown in Fig. \ref{fig:data_efficiency}, in the experimental groups where $D_{fused}=4$ and $D_{fused}=32$, the 4-qubit VQC model, denoted as the $N_q=4$ curve, demonstrated classification accuracy comparable to that of the classical MLP. Although this VQC model contains fewer parameters than the MLP, the results show it can effectively fit the data distribution. This indicates that the VQC-based lightweight classifier can serve as an effective alternative to classical fully connected networks, maintaining comparable detection performance while reducing the number of model parameters.

In the results for $D_{fused}=16$, we observed that the VQC models with 4 and 6 qubits exhibited a degree of fluctuation in ACER at low data ratios, rising to approximately 1.0\%. However, when the number of qubits was increased to 8 or 12, the model performance became more stable. This observation implies a dependency between the number of qubits in the VQC and the dimensionality of the input features. When using angle encoding strategies, an insufficient number of qubits may limit the circuit's ability to encode high-dimensional features. Appropriately increasing the number of qubits helps improve the expressive capacity of the circuit, thereby enhancing classifier stability at specific dimensions.

Regarding data dependency, the experiment compared model convergence under varying training data amounts. In the settings of $D_{fused}=8$ and $D_{fused}=32$, VQCs with 8 or 12 qubits maintained low ACER values even when the training data was reduced to 10\%. This demonstrates that the VQC architecture can achieve convergence with limited sample sizes, indicating potential for few-shot learning. This characteristic makes it suitable for face anti-spoofing tasks where large-scale attack samples are difficult to obtain.

\section{Conclusion and Discussion}
\label{sec:conclusion}

In summary, this study proposes and evaluates a hybrid MPS-VQC framework designed to address the challenge of efficiently encoding high-dimensional multimodal data into limited quantum resources in face anti-spoofing tasks. By systematically bridging classical visual features with quantum states, the proposed framework establishes a resource-efficient pathway for deploying quantum machine learning on NISQ devices.

Experimental results indicate that the MPS-based feature fusion-compression module effectively achieves structured compression of high-dimensional multimodal data. The MPS simulates the coarse-graining process of the renormalization group flow, substantially reducing feature dimensionality while preserving key discriminative information required to distinguish between live and spoof samples. This structured dimensionality reduction avoids the information loss typically associated with traditional linear methods under extreme compression ratios.

Furthermore, the introduction of a nonlinear activation mechanism significantly enhances training stability. In processing high-dimensional and long-sequence features, this mechanism acts similarly to a classical gating unit, smoothing the optimization landscape and suppressing gradient oscillation phenomena in long tensor chains. This ensures numerical robustness and convergence speed when handling complex multimodal distributions.

Crucially, this work validates the feasibility of achieving high-performance quantum classification under resource constraints. Leveraging the efficient preprocessing of the MPS frontend, the architecture achieves detection accuracy comparable to classical baselines like ResNet with fewer parameters. This demonstrates that appropriate feature engineering can circumvent the optimization difficulties of direct variational search in large Hilbert spaces, providing theoretical support for high-fidelity mapping from classical data to quantum circuits.

Future work will focus on deploying this architecture on real noisy quantum processors to evaluate the impact of quantum noise on the MPS entanglement structure. Additionally, we plan to explore two-dimensional or hierarchical tensor network structures to investigate their advantages in processing spatially structured data, and to conduct interpretability studies based on quantum entanglement entropy to analyze correlation mechanisms among multimodal features.



\section*{Funding}
This work was supported by the CPS-Yangtze Delta Region Industrial Innovation Center of Quantum and Information Technology-MindSpore Quantum Open Fund.

\section*{Code Availability}
The implementation code is available at: https://github.com/arctenio/MPS-VQC.

\section*{Acknowledgments}
The authors acknowledge the MindSpore Quantum team for their technical support and the open-source community for providing the simulation framework.

\bibliography{bibliography}

\end{document}